\documentclass[sigconf,  screen, authordraft,review=false]{acmart} 

\AtBeginDocument{%
  }

\usepackage{algorithm}
\usepackage{algorithmic}
\usepackage{listings}

\definecolor{juliagreen}{rgb}{0.22,0.596,0.149}
\definecolor{juliapink}{rgb}{0.584,0.345,0.698}
\lstdefinelanguage{Julia}%
  {morekeywords={abstract,break,case,catch,const,continue,do,else,elseif,%
      end,export,false,for,function,immutable,import,importall,if,in,%
      macro,module,otherwise,quote,return,switch,true,try,type,typealias,%
       using,while},%
        morekeywords=[2]{@kernel,@synchronize, @index, @private, @localmem},%
        morekeywords=[3]{@unroll},%
   sensitive=true,%
   alsoother={\$},%
   morecomment=[l]{\#},%
   morecomment=[n]{\#=}{=\#},%
   morestring=[s]{"}{"},%
   morestring=[m]{'}{'},%
}[keywords,comments,strings]%

\copyrightyear{2025}
\acmYear{2025}
\setcopyright{rightsretained}
\acmConference[ICPP '25]{54th International Conference on Parallel Processing}{September 08--11, 2025}{San Diego, CA, USA}
\acmBooktitle{54th International Conference on Parallel Processing (ICPP '25), September 08--11, 2025, San Diego, CA, USA}

\acmDOI{10.1145/3754598.3754667}
\acmISBN{979-8-4007-2074-1/25/09}

\begin{document}

\title{Performant Unified GPU Kernels for Portable Singular Value Computation Across Hardware and Precision}

\author{Evelyne Ringoot}
\affiliation{%
  \institution{Massachusetts Institute of Technology}
  \city{Cambridge, MA}
  \country{USA}}
\email{eringoot@mit.edu}
\orcid{}

\author{Rabab Alomairy}
\affiliation{%
  \institution{Massachusetts Institute of Technology}
  \city{Cambridge, MA}
  \country{USA}}
\email{rabab.alomairy@mit.edu}
\orcid{}

\author{Valentin Churavy}
\affiliation{%
  \institution{University of Mainz \& University of Augsburg}
  \city{Mainz}
  \country{Germany}}
\email{vchuravy@uni-mainz.de}
\orcid{0000-0002-9033-165X}

\author{Alan Edelman}
\affiliation{%
  \institution{Massachusetts Institute of Technology}
  \city{Cambridge, MA}
  \country{USA}}
\email{edelman@mit.edu}
\orcid{}

\renewcommand{\shortauthors}{Ringoot, Alomairy, Churavy and Edelman}

\begin{abstract}
This paper presents a portable, GPU-accelerated implementation of a QR-based singular value computation algorithm in Julia. The singular value decomposition (SVD) is a fundamental numerical tool in scientific computing and machine learning, providing optimal low-rank matrix approximations.
Its importance has increased even more in large-scale machine learning pipelines, including large language models (LLMs), where it enables 
low-rank adaptation (LoRA).
The implemented algorithm is based on the classic two-stage QR reduction, consisting of successive matrix reduction to band form and bidiagonal form. 
Our implementation leverages Julia’s multiple dispatch and metaprogramming capabilities, integrating with the GPUArrays and KernelAbstractions frameworks to provide a unified type and hardware-agnostic function. It supports diverse GPU architectures and data types, and is, to our knowledge, the first GPU-accelerated singular value implementation to support Apple Metal GPUs and half precision.  Performance results on multiple GPU backends and data types demonstrate that portability does not require sacrificing performance: the unified function outperforms most linear algebra libraries (MAGMA, SLATE, rocSOLVER, oneMKL) for matrix sizes larger than $1024 \times 1024$, and achieves 80\%-90\% of the performance of cuSOLVER for large matrices.

\end{abstract}



\maketitle
\vspace{-0.5cm}
© {Ringoot, Alomairy, Churavy, Edelman | ACM} {2025}. 
This is the author's pre-print version of the work. It is posted here for your personal use. The definitive Version of Record is to appear in ICPP'25, https://doi.org/10.1145/3754598.3754667.

\section{Introduction}
The singular value decomposition (SVD) is a cornerstone of numerical linear algebra. 
Its ability to compute optimal low-rank approximations of matrices makes it essential for tasks such as dimensionality reduction, 
scientific computing\cite{doi:10.1137/21M1445934}, signal processing\cite{svd_imageprocessing}, quantum information theory\cite{svd_quantuminfo}, and low-rank adaptations in large language models (LLMs)~\cite{sun2025transformersquared,wang2025dobisvddifferentiablesvdllm, li2025adasvdadaptivesingularvalue}. As machine learning workloads continue to grow in scale and new AI accelerators emerge, there is an increasing demand for portable, highly performant SVD implementations that can run efficiently across diverse hardware platforms, including modern GPUs and specialized AI processors.

Several libraries support SVD computation, either CPU-resident like LAPACK~\cite{anderson1999lapack}, GPU-accelerated like SLATE~\cite{gates2025evolution}, MAGMA~\cite{abdelfattah2024magma} and rocSOLVER~\cite{rocsolver} or GPU-resident like cuSOLVER~\cite{cusolver}. Yet, one limitation most libraries have in common is their reliance on low-level vendor-specific kernels, requiring different implementations for each hardware backend, and as a result, supporting only a limited number of hardware vendors. Notably, Apple Metal GPUs are not supported by any of the packages, and their built-in Metal Performance Shaders (MPS) framework~\cite{applemps} does not support singular value computations. Similarly, separate optimized implementations are typically included per input data precision, increasing development time and difficulty of integration of new data precision formats, and providing limited support for type-agnostic or low-precision arithmetic — a growing need in modern AI and inference workloads, where mixed and low-precision arithmetic are key to maximizing performance while minimizing memory usage and energy efficiency.

To address these limitations, we present a performant, portable GPU-accelerated implementation of two-stage QR-based singular value computation developed as open-source software using abstraction layers in the Julia language. Our approach employs a two-stage QR reduction scheme~\cite{10.1145/2063384.2063394} while exposing a single, unified API that spans multiple GPU architectures and input data precisions. By leveraging Julia’s multiple dispatch and dynamic type inference, our implementation generates architecture- and precision-specific LLVM machine code at compile time. GPU execution is abstracted through the GPUArrays.jl\cite{gpuajl} and KernelAbstractions.jl\cite{kajl} frameworks, enabling backends for NVIDIA, AMD, Intel, and Apple GPUs within a common interface.
The original contribution of this work can be summarized  as follows:
\begin{enumerate}

\item We present a unified, device-agnostic implementation that achieves high performance on NVIDIA, AMD, Intel, and Apple GPUs, eliminating the need for architecture-specific code duplication. This is the first GPU-accelerated singular value solver portable to Apple Metal, filling a critical gap left by existing vendor-specific libraries.

\item We introduce a type-agnostic singular value implementation for computing singular values using a QR-based decomposition. To the best of our knowledge, this is the first GPU-based SVD implementation to support half-precision (FP16), enabling low-memory, high-throughput computations for AI workloads.

\item We benchmark our implementation against state-of-the-art libraries, including SLATE, MAGMA, oneMKL, cuSOLVER, and rocSOLVER. For large matrix sizes, our unified API outperforms all of them, except cuSOLVER, where we achieve 80–90\% of its performance on high-performance NVIDIA GPUs at large matrix sizes. 

{\item We present a GPU parallelization strategy for computing singular values that outperforms existing open-source libraries (e.g., rocSOLVER). It approaches the performance of proprietary software such as cuSOLVER. Our implementation is released as open-source software\cite{the-code}, enabling reuse either through direct integration into abstraction-based frameworks or by linking it as a standalone library in larger code-bases\cite{embeddingjulia}.}

\item We demonstrate that by abstracting the appropriate hyperparameters in GPU kernels, performance portability can be achieved without reimplementing the full algorithm through hyperparameter tuning. 
\end{enumerate}
\section{Related Work}\label{sect:lit}
Hardware-specific data precision-optimized dense linear algebra libraries have been extensively developed over the past half-century (Section \ref{sect:denselit}).  Motivated by the emergence of heterogeneous computing clusters with different types of GPUs available from several hardware vendors\cite{10820782, 9652859,10196600}, the scientific community has recently displayed significant interest in developing portable abstraction layers (Section \ref{sect:abstractlit}). In this work, the Julia abstractions frameworks enable us to develop a unified hardware- and type-adaptive function for calculating singular values (Section \ref{KAGPUA}).

\subsection{Dense Linear Algebra Libraries}\label{sect:denselit}

Classical libraries such as LAPACK~\cite{anderson1999lapack} and ScaLAPACK~\cite{blackford1997scalapack} provide highly scalable implementations of SVD for CPUs, targeting shared- and distributed-memory systems. More recently, the PLASMA and DPLASMA libraries\cite{10.1145/3264491, 6008998} have been optimized for shared memory systems but do not support singular value calculations. However, none of these libraries is designed for modern GPU architectures. As such, they are limited in their ability to leverage the massive parallelism and high memory bandwidth offered by accelerators. Several GPU-focused libraries have emerged to exploit the massive parallelism and memory bandwidth available on modern accelerators. Notably, MAGMA~\cite{abdelfattah2024magma} provides GPU-accelerated hybrid CPU-GPU versions of LAPACK routines, including SVD, using hybrid CPU-GPU algorithms by offloading compute-intensive tasks into GPUs. 
Meanwhile, the SLATE~\cite{gates2025evolution, alomairy2022communication} library has emerged as a modern C++ library designed to replace ScaLAPACK by providing multicore distributed-memory linear algebra operations optimized for heterogeneous CPU-GPU architectures. Both SLATE and MAGMA support multiple vendor backends (e.g., NVIDIA, AMD, Intel), but rely on low-level, vendor-specific kernels and expose backend-dependent APIs.

In contrast to the hybrid multi-architecture solvers, cuSOLVER, part of NVIDIA’s CUDA toolkit, offers a GPU-only implementation of SVD and related dense linear algebra routines on Nvidia hardware.\cite{cusolver} Similarly, rocSOLVER, developed by AMD as part of the ROCm software stack, provides hybrid CPU-GPU routines targeting AMD GPUs.\cite{rocsolver}
Intel’s oneAPI initiative provides linear algebra support through oneMKL via DPC++/SYCL. Meanwhile, Apple’s Metal Performance Shaders (MPS) framework offers accelerated linear algebra routines for Apple GPUs, but support for advanced factorizations such as SVD is absent, and documentation remains limited~\cite{applemps}. 

\subsection{Abstraction Frameworks}\label{sect:abstractlit}

Linear algebra libraries that are not easily portable to novel architectures pose a burden in terms of development and maintenance, and limit their use in heterogeneous computing environments. Several frameworks have been created that aim to be both performant and portable to different GPU hardware, the most notable being:
\begin{itemize}
    \item OpenCL~\cite{stone2010opencl} providing low-level C-like portability over NVIDIA and AMD
    \item SyCL~\cite{10.1145/2791321.2791345, 10.1145/3529538.3530005} leveraging templates in C++ to provide portability over AMD, NVIDIA and Intel
    \item Kokkos, which provides abstractions and linear algebra functionality for AMD, NVIDIA, and Intel in C++~\cite{davis2024evaluative, 9485033}
    \item HIP framework, which aims to provide portability between CUDA and ROCm kernels~\cite{https://doi.org/10.1002/cpe.7866}
    \item Julia, which provides general abstractions over all levels of software programming ~\cite{besard2019rapid, 8471188, 10.1145/3276490}
\end{itemize}
Many other efforts are ongoing~\cite{https://doi.org/10.1002/cpe.4117, 10883176, 8945642}, highlighting the growing interest in abstractions for GPU kernel programming. Similarly, other aspects of performant large-scale linear algebra have been abstracted, for example the BLIS and FLAME libraries that aim to abstract many linear algebra algorithms as compositions of the most simple BLAS operations~\cite{IGUAL20121134, 10.1145/2764454}, the Chameleon/DPLASMA~\cite{faverge2023programming, bosilca2011flexible} and StarPU/PARSEC~\cite{10.1007/978-3-642-03869-3_80, cao2022framework} abstractions for runtime scheduling, or AL4SAN for abstracting several runtime schedulers~\cite{alomairy2020abstraction}.
In this work, we opt to use the Julia programming framework as it combines linear algebra abstractions, GPU kernel abstractions, data type abstractions, and scheduling abstractions in a single high-level framework where the metaprogramming capabilities and just-in-time type-inference facilitate the development of high-level abstraction layers. 
In particular, the utility of the Julia language Array framework for generic linear algebra abstractions with minimal code length has been demonstrated through the development of the NextLA.jl library, which targets a type- and hardware-agnostic API for dense linear algebra~\cite{carrica2025toward,dla} across CPUs and GPUs. This unified API reduces development time and enhances composability: new data types can be supported by the existing implementations without modifications, making the library future-proof and suitable for heterogeneous environments.

\subsection{Compiler-Enabled Specialized Performance}\label{KAGPUA}

This work proposes a unified function over input data precision and GPU hardware vendor.  Through Julia's multiple-dispatch and type-inference~\cite{10.1145/3276490}, it is possible to write an abstract function targeting several data types, which at compile-time get inferred 'just-ahead-of-time' through LLVM and optimized for the input data type to achieve comparable performance to specialized functions. For example, Algorithm \ref{alg:stageone} is the high-level user interface that has as a parameter the hardware backend and the data precision {\tt T}. At compile-time, based on user inputs, the function will be specialized based on the input data types.
The {\tt GPUArrays} abstraction~\cite{gpuajl, 8471188} provides a high-level interface for high-performance numerical computing on GPUs that targets a wide range of hardware backends. The framework utilizes type-inference capabilities to replace part of the code generation process so that the compiler generates accelerator-optimized Intermediate Representations (e.g., NVPTX for CUDA and NVIDIA hardware) for several GPU backends, while also providing generic functionality, once again providing performance comparable to specialized functions through a generic interface. {To execute more complex linear algebra operations, \texttt{GPUArrays.jl} relies on vendor-provided libraries. Specifically, \texttt{svdvals} defaults to \texttt{cuSOLVER} on NVIDIA hardware and \texttt{rocSOLVER} on AMD hardware, while no implementation is currently available for Apple GPUs.} {\tt KernelAbstractions.jl} is a specialized interface providing access to kernel functionality across hardware vendors ({\tt CUDA.jl}~\cite{8471188}, {\tt AMDGPU.jl}~\cite{AMDGPU}, {\tt oneAPI.jl}~\cite{oneapi}, {\tt Metal.jl}~\cite{metal}), compiling the generic kernel interface to hardware-specific Intermediate Representations at compile time. This toolkit allows us to program GPU kernels in a single interface and provides a single interface to the end-user that provides portable performance across GPU architectures. Earlier work leveraging {\tt KernelAbstractions.jl} demonstrated the viability of the approach for scientific machine learning~\cite{UTKARSH2024116591}, and earlier work on \texttt{NextLA.jl} demonstrated the viability of generic abstractions for dense linear algebra across CPUs and GPUs.~\cite{carrica2025toward,dla} 

\section{Technical Background and Implementation}\label{sect:methods}

The singular value decomposition of a square matrix $A\in \mathbb{C}^{n \times n}$ is the matrix factorization:
\begin{equation}
    A = U \Sigma V^T,
\end{equation}

where $U,V \in \mathbb{C}^{n \times n}$ are unitary matrices of left and right singular vectors and $\Sigma=\text{diag}(\sigma_1,...,\sigma_n)$ is a diagonal matrix of singular values. 
Typical approaches for computing all singular values of dense matrices include Jacobi-based methods~\cite{10.1145/3503221.3508443}, divide-and-conquer methods~\cite{GATES20183}, and QR-based methods~\cite{BUTTARI200938}. Specialized algorithms exist for non-square or dense matrices. 
In this work, we adopt the classical two-stage QR approach introduced by Haidar et al.~\cite{10.1145/2063384.2063394}, which has been implemented in state-of-the-art libraries such as MAGMA~\cite{Dongarra2014} and SLATE~\cite{gates2020slate} for CPU-GPU hybrid platforms. {This approach consists of reducing a dense matrix to diagonal form by successive orthogonal transformations: a compute-bound reduction to band form in the first phase and a further reduction to bidiagonal and diagonal form in the second phase. \label{sect:phase2} The first phase, the focus of this paper, is summarized in section \ref{alg:bired}, and the design of the implemented GPU kernels, a custom parallelization of the classical householder QR factorization for GPU architectures, is discussed in Section \ref{sect:gpukernel}. 
The second, memory-bound stage is performed through a GPU-accelerated band-to-bidiagonal reduction, following the cache-efficient tile kernels proposed by Haidar et al.~\cite{6267821} and adopting a communication-avoiding strategy inspired by Ballard et al.~\cite{10.1145/2145816.2145822} on the GPU. 
{A detailed study of the reduction from band form to bidiagonal form is the topic of future work.}
The final stage -- reduction from bidiagonal form to singular values -- is the least computationally expensive among the three~\cite{doi:10.1137/17M1117732}. For this step, we rely on existing high-quality CPU implementations, specifically LAPACK’s divide-and-conquer algorithm~\cite{69824}. The benchmarking strategy used is discussed in Section \ref{sect:benchmark}. 

\subsection{Phase 1: reduction to band form}\label{alg:bired}
\begin{figure}[ht]
  \centering
  \includegraphics[width=\linewidth]{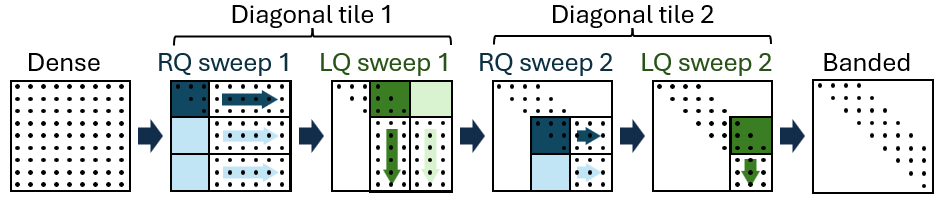} 
  \caption{ Phase 1 algorithm (see Algorithm~\ref{alg:redtoband}). For each diagonal tile, an RQ sweep applies an orthogonal transformation that makes the tile upper triangular and annihilates the column below. Subsequently, an LQ sweep transforms the tile to the right of the diagonal into a lower triangular form and annihilates the row to its right.
}\label{fig:brdalgfig}
  \Description{Six matrices next to each other with an arrow between each, the first is dense, the second and third are labeled 'diagonal tile 1', with the second being RQ sweep and the third LQ sweep. The second matrix, RQ sweep 1, has three times three tiles. The top left diagonal tile is upper triangular, and the two tiles below are empty. The third matrix, LQ sweep 1, has the same structure for the first column, and has the top middle tile lower triangular and the top right tile empty in addition. The fourth and fifth matrices are labeled 'diagonal tile 2', with the fourth being the RQ sweep and the fifth the LQ sweep. The fourth matrix has the same top row and first column as the third, with, in addition, the middle tile, the second diagonal tile, the upper triangular, and the bottom middle tile empty. The fifth matrix is similar, with, in addition, the right middle tile lower diagonal. The sixth matrix shows the result: a banded matrix, with 4 entries per row: the diagonal and three entries next to the diagonal.  }
  \vspace{-0.3cm}
\end{figure}

Figure \ref{fig:brdalgfig} schematically represents the well-known phase 1 algorithm~\cite{10.1145/2063384.2063394}, also detailed in Algorithm \ref{alg:redtoband}. 
We follow the standard LAPACK naming convention, accompanied by concise explanations. For comprehensive details, see~\cite{doi:10.1137/17M1117732}. For each diagonal tile (line 1), a block Householder reflector is computed to transform the tile into upper triangular form (line 2). This transformation is then applied to the entire row in a trailing submatrix update (line 3). Below-diagonal tiles are successively annihilated by jointly computing their QR decomposition with the diagonal block—this panel factorization step is shown in line 4. The resulting orthogonal transformation zeroes out the below-diagonal tiles, followed by updates to both the tile row and the first row of the trailing submatrix (line 6). This completes the RQ sweep, as illustrated in the second matrix of Figure~\ref{fig:brdalgfig}. Next, the transpose of this algorithm is applied to the tile to the right of the diagonal tile. It transforms the tile into a lower triangular form and annihilates the tiles to its right through an LQ panel factorization and trailing submatrix update (lines 8–13 in Algorithm 1, shown as “LQ sweep 1” in the third matrix of Figure~\ref{fig:brdalgfig}). The RQ and LQ sweeps are repeated for each diagonal tile (see matrices 4 and 5 in Figure 1), progressively reducing the matrix to a band form. This concludes Phase 1 of the algorithm.

} 

 While the original algorithm utilizes separate kernels for QR and LQ decompositions, our implementation performs only QR decomposition and computes LQ decomposition by applying QR to the transpose of the input tiles, as outlined in Algorithm~\ref{alg:redtoband}. To avoid the overhead of explicit data transposition, we leverage Julia’s lazy transpose operator, which enables index-level transposition without altering the underlying memory layout~\cite{churavy2019transparent}. In particular, we implement lines 2–7 of Algorithm~\ref{alg:redtoband}—corresponding to the panel factorization and trailing submatrix update—as a single function, \texttt{GETSMQRT}. This function is invoked in Algorithm~\ref{alg:stageone}, which performs the stage one reduction to band form. In line 5 of Algorithm~\ref{alg:stageone}, a lazy transpose is applied, enabling reuse of the same function for both QR and LQ operations.

\begin{algorithm}
\small
\caption{SVD stage 1: reduction to band form \cite{10.1145/2063384.2063394} }\label{alg:redtoband}
\begin{algorithmic}[1]
\FOR{Diagonal tile $k$ : 1 $\rightarrow$ N }
\STATE{GEQRT(Tile$_{k,k}$)} \qquad \qquad \;  \COMMENT{Calculate Householder QR}
\STATE{UNMQR(Row$_k$)} \qquad \qquad \;  \COMMENT{Apply Householder QR}
\FOR{TileRow $l$ below $k$ : k+1 $\rightarrow$ N }
\STATE{TSQRT(Tile$_{k,k}$, Tile$_{l,k}$)} \quad \COMMENT{Calculate Householder QR}
\STATE{TSMQR(Row$_k$, Row$_l$)} \quad \; \COMMENT{Apply Householder QR}
\ENDFOR
\STATE{GEQRT(Tile$_{k,k+1}^T$)} \qquad \; \qquad  \COMMENT{Calculate Householder QR}
\STATE{UNMQR(Col$_{k+1}^T$)} \qquad \qquad \; \; \COMMENT{Apply Householder QR}
\FOR{TileCol $l$ right of $k+1$ : k+2 $\rightarrow$ N }
\STATE{TSQRT(Tile$_{k,k+1}^T$, Tile$_{k,l}^T$)} \quad \COMMENT{Calculate Householder QR}
\STATE{TSMQR(Col$_{k+1}^T$, Col$_l^T$)} \quad \; \: \COMMENT{Apply Householder QR}
\ENDFOR
\ENDFOR
\end{algorithmic}

\end{algorithm}

\begin{algorithm}
\scriptsize
\caption{Julia code: reduction to band form}\label{alg:stageone}
\begin{lstlisting}[language=Julia]
function banddiag!(A::AbstractGPUMatrix{T}, Tau::AbstractGPUMatrix{T}, nbtiles::Int, backend) where T
    for k in 1:(nbtiles-1)
        GETSMQRT!(A,Tau,k, nbtiles, backend)
        GETSMQRT!(A',Tau,k, nbtiles, backend, LQ=true)
    end
    GETSMQRT!(A,Tau,nbtiles, nbtiles, backend)
end
\end{lstlisting}
\vspace{-0.2cm}
\end{algorithm}

\subsection{Performant GPU Kernel Design for singular values}\label{sect:gpukernel}
Algorithm \ref{alg:redtoband} is a restructured from the classic householder QR (which would operate row-by-row instead of tile-by-tile) to reduce the serial bottlenecks (panel factorizations), and leverage more BLAS3 operations for parallelism (trailing submatrix updates)\cite{10.1145/2063384.2063394}. In this work, we develop custom highly performant general GPU kernels for both stages.
{
\paragraph{Panel Factorization.}
Even after restructuring the algorithm, the panel QR factorization (\texttt{GEQRT}/\texttt{TSQRT}) remains a serial bottleneck due to critical sections lacking parallelism and increased communication cost at high occupancy. To mitigate this, the implementation assigns a single thread block (or workgroup) of \texttt{TILESIZE} threads, aiming to minimize communication overhead, as shown in Algorithm~\ref{alg:geqrt}---a custom parallel version of the classical Householder QR algorithm for \texttt{GEQRT}. This approach can be extended to \texttt{TSQRT} by including a second tile. Algorithm~\ref{alg:geqrt} performs an in-place QR factorization of tile $A$, storing the corresponding $\tau$ coefficients (stored as normalized $\hat{\tau}$ such that the householder reflector $H=I- \hat{\tau} \cdot  \mathbf{ v  v^T} $). Each thread uses thread-local memory (registers, line 2) to store one column $A_i$ of the tile (line 4). For each Householder vector to be computed (i.e., for each column, line 5), one column of size \texttt{TS} is loaded into shared thread-block memory (line 3), where column $A_k$ is saved (line 6), and its squared norm is computed and shared across the block (line 7). Each thread then accesses this shared column to compute the first element of the Householder vector $x$ (line 10-11), the $\tau$ for the current reflector (line 12),  and the vector product of the thread-block column with the thread-local column (line 13), with a correction step for small reflectors (line 14--15, $\epsilon$ is machine precision for a given precision). This is followed by an update of the remaining tile columns (lines 16--18). Threads collaboratively write back one row of the tile after each iteration (lines 19, 22), and the complete set of $\hat{\tau}$ values after the factorization completes (line 23). We follow the Julia [row, column] convention in the notation of the matrix partitions.

\begin{algorithm}[ht]
\small
\caption{GPU kernel GEQRT: 1 thread block with TILESIZE (TS) threads }\label{alg:geqrt}
\begin{algorithmic}[1]
\STATE{{\bf INPUT/ OUTPUT}: A (TS$\times$TS), $\hat{\tau}$ (TS)}
\STATE{{\bf Thread memory}: $A_i$ (TS), $\hat{\tau}_i$ (1), $\rho$ (1), $\rho'$ (1) , $x$ (1)}
\STATE{{\bf Block memory}: $A_k$ (TS), $|A_k|^2$ (1) }
\STATE{$\forall$ Thread $i$: $A_i \gets A[:,i] \qquad$   }
\FOR{ $k$ : 1 $\rightarrow$ TS-1 }
\STATE{Thread $i=k$: $A_k \gets A_i$}
\STATE{ \qquad \qquad  \quad {$|A_k|^2$} $\gets  ||A_i$[k+1:]$||^2$  }
\STATE{synchronize threads}
\STATE{$\forall$ Thread $i\geq k$: $\rho \gets A_i$[k+1:]$^T\cdot A_k$[k+1:]    }
\STATE{ \qquad \qquad  \qquad \; $x = A_k[k] - \sqrt{A_k[k]^2+|A_k|^2} $ } \quad \COMMENT{if $A_k[k]<0$}
\STATE{ \qquad \qquad  \qquad \quad \; $  = A_k[k] + \sqrt{A_k[k]^2+|A_k|^2} $ } \quad \COMMENT{if $A_k[k]\geq 0$}
\STATE{ \qquad \qquad  \qquad \;  $\hat{\tau}_i \gets 2 x^2 /  ( x^2+ |A_k|^2) $  }
\STATE{ \qquad \qquad  \qquad \; $\rho'  \gets  (\hat{\tau}_i / x) \cdot (A_i[k]\cdot x+\rho) $}
\STATE{ \qquad \qquad  \qquad \; \textbf{if $|x|<10\epsilon$: $x \gets 10\epsilon, \hat{\tau}_i \gets 2$}
\STATE{ \qquad \qquad  \qquad \qquad \qquad \qquad $\rho'  \gets 2 \cdot (A_i[k]+\rho/x)   $ }  }
\STATE{ \qquad \qquad  \qquad \; $A_i[k] \gets A_i[k] - \rho' $ }
\STATE{$\forall $ Thread $i>k$: $A_i$[k+1:]$ \gets A_i$[k+1:]$ - \rho' \cdot (A_k$[k+1:]$/x)  $ }
\STATE{Thread $i=k$: $A_i$[k+1:]$ \gets A_i$[k+1:]$/x $ }
\STATE{$\forall$ Thread $i$: A[k,i] $ \gets A_i[k]$   }
\STATE{synchronize threads}
\ENDFOR
\STATE{$\forall$ Thread $i$: A[TS,i] $ \gets A_i[TS]$   }
\STATE{ \qquad \qquad  \quad $\hat{\tau}$[i] $\gets$ $\hat{\tau}_i$  }
\end{algorithmic} 
\end{algorithm}

\paragraph{Register Pressure and SPLITK Parallelism.}
While Algorithm~\ref{alg:geqrt} is highly sensitive to register pressure, the \texttt{GEQRT} kernel remains inherently limited by serial communication bottlenecks. As only a single thread block is launched, the goal of the algorithm is to maximize L1 cache utilization rather than occupancy. Register usage can be reduced by fine-tuning the \texttt{TILESIZE} parameter. The optimal value maximizes L1 usage while minimizing register spills and is typically architecture dependent. Algorithm~\ref{alg:geqrt} can be trivially extended to use \texttt{SPLITK}~$\times$~\texttt{TILESIZE} threads by dividing each tile column among \texttt{SPLITK} threads and performing intermediate reductions through shared memory. This \textit{split-K} strategy increases occupancy but introduces additional inter-thread communication, which can be optimized based on the hardware architecture. It is important to note that \texttt{TILESIZE} is an algorithmic parameter: it controls the number of iterations in the \texttt{for} loops of Algorithm~\ref{alg:stageone}, thereby influencing the dependency graph. In contrast, \texttt{SPLITK} is a purely computational parameter—the same operations are executed in the same order, either in parallel using \texttt{SPLITK}$=8$ threads, or serially using \texttt{SPLITK}$=1$ thread.

\paragraph{Trailing Submatrix Update.}
The trailing submatrix update (\texttt{UNMQR} / \texttt{TSMQR}) applies the Householder vectors computed in \texttt{GEQRT}/\texttt{TSQRT} to the remaining rows. This is a massively parallel \texttt{BLAS3}-level operation, as each column can be updated independently using the Householder reflectors. Algorithm~\ref{alg:unmqr} presents the custom parallelization strategy for \texttt{UNMQR}, in which a group of columns, specified by the \texttt{COLPERBLOCK} parameter, is processed by a single workgroup (thread block). \texttt{COLPERBLOCK} is a purely computational hyperparameter that can be tuned for specific hardware: smaller values reduce register pressure, while larger values mitigate warp divergence and reduce the number of memory accesses. The algorithm performs an in-place update of tile $X$, using the matrix $A$ and vector $\hat{\tau}$ produced by \texttt{GEQRT}. Each thread utilizes thread-local memory (registers, line 3) to store one column $X_i$ of tile $X$ (line 5), while the $\hat{\tau}$ values are cooperatively loaded into shared memory by all threads—i.e., each thread may load one or more elements, depending on the \texttt{COLPERBLOCK} setting. For every Householder vector to be applied (i.e., each column of $A$, line 7), column $A_k$ is cooperatively loaded into shared memory (line 8). Each thread then computes its partial vector product with $X_i$, scaled by $\hat{\tau}[k]$ (line 10), and applies the Householder transformation (line 11--12). After processing all reflectors, the updated columns $X_i$ are written back to global memory (line 15).

\begin{algorithm}[ht]
\small
\caption{GPU kernel UNMQR: N thread blocks with COLPERBLOCK (CPB) threads \raggedright }\label{alg:unmqr}
\begin{algorithmic}[1]
\STATE{{\bf INPUT}: A (TS$\times$TS), $\hat{\tau}$ (TS)}
\STATE{{\bf INPUT/ OUTPUT}: X ($TS \times \; N\cdot CPB $)}
\STATE{{\bf Thread memory}: $X_i$ (TS), $\rho$ (1)}
\STATE{{\bf Block memory}: $A_k$ (TS), $\hat{\tau}'$ (TS) }
\STATE{$\forall$ Threads in block cooperatively: $\hat{\tau}' \gets \hat{\tau}  \qquad$   }
\STATE{$\forall$ Thread $i$ in block $n$: $X_i \gets X$[:,$n\cdot CPB\;+$i] \qquad  }
\FOR{ $k$ : 1 $\rightarrow$ TS-1 }
\STATE{ $\forall$ Threads in block cooperatively: $A_k \gets A$[:,k] }
\STATE{synchronize threads}
\STATE{$\forall$ Thread $i: \rho  \gets \hat{\tau}'[k] \cdot (X_i\text{[k]}+X_i\text{[k+1:]}^T\cdot A_k\text{[k+1:]})$}
\STATE{ \qquad \qquad  \quad $X_i\text{[k:]} \gets X_i\text{[k:]} - \rho $ }
\STATE{ \qquad \qquad  \quad $X_i\text{[k+1:]} \gets X_i\text{[k+1:]} - \rho \cdot   A_k\text{[k+1:]} $ }
\STATE{synchronize threads}
\ENDFOR
\STATE{$\forall$ Thread $i$ in block $n$: $ X$[:,$n\cdot CPB\;+$i] $\gets X_i$ \qquad  }
\end{algorithmic}
\end{algorithm}
The algorithm can be trivially extended to \texttt{TSMQR} by including a second tile. However, extension to \texttt{SPLITK} threads per column is not applied for this kernel. As the matrix size increases, the number of independent columns becomes sufficiently large to achieve full occupancy. In this scenario, introducing a split-K computation would only increase communication cost without providing further occupancy gains. This behavior was confirmed through empirical evaluation on large matrices.
\begin{algorithm}[ht]
\caption{Julia code: Fused TSMQR Kernel}\label{alg:kakernel}
\begin{lstlisting}[language=Julia]
@kernel function TSMQR_fused_kernel!(X, Y, @Const(A), 
                            @Const(T), nbrows::Int)
    g = @index(Group, Linear)
    i = @index(Local, Linear)
    Xi = @private eltype(A) (TILESIZE)
    Yi = @private eltype(A) (TILESIZE)
    Ak= @localmem eltype(A) (TILESIZE)
    Tk = @localmem eltype(A) (TILESIZE)
    
    @unroll for j in 1:TILESIZE
        Yi[j] = Y[...] 
    end
    for _ in 1:nbrows
        @unroll for j in 1:TILESIZE
            Xi[j] = X[...] 
        end
        @unroll for j in 0:(TILESIZE/COLPERBLOCK)-1
            Tk[j*COLPERBLOCK+i]=T[...]
        end 
        for k in 1:TILESIZE
            @unroll for j in 0:(TILESIZE/COLPERBLOCK)-1
                Ak[j*COLPERBLOCK+i]=A[...]
            end 
            @synchronize  
            Xik= zero(eltype(A))
            @unroll for j in 1:TILESIZE
                Xik += Ak[j] * Xi[j]
            end
            Xik= (Xik+Yi[k])* Tk[k]
            Yi[k] -= Xik
            @unroll for j in 1:TILESIZE
                Xi[j] -= Xik * Ak[j]
            end
            @synchronize  
        end
        @unroll for j in 1:TILESIZE
            X[...] = Xi[j]
        end
    end
    @unroll for j in 1:TILESIZE
        Y[...] = Yi[j]
    end
end
\end{lstlisting}
\vspace{-0.2cm}
\end{algorithm}
\paragraph{Fusing TSQRT and TSMQR Kernels} In Algorithm~\ref{alg:redtoband}, the loops on lines 4–7 and 10–13 perform, for each tile row, a TSQRT (panel factorization) followed by a TSMQR (trailing submatrix update). While this computation is originally structured to apply TSQRT and TSMQR sequentially along tile rows, it can be reordered such that all TSQRT operations are performed first across all tile rows, followed by all TSMQR operations. The successive \texttt{TSQRT} and \texttt{TSMQR} operations can be merged into fused kernels, \texttt{FTSQRT} and \texttt{FTSMQR}, respectively. These fused kernels process the entire panel and apply the corresponding reflectors in a single kernel launch, rather than issuing one launch per row, as illustrated in Figure~\ref{fig:fuse}. 
\begin{figure}[ht]
  \centering
  \includegraphics[width=0.65\linewidth]{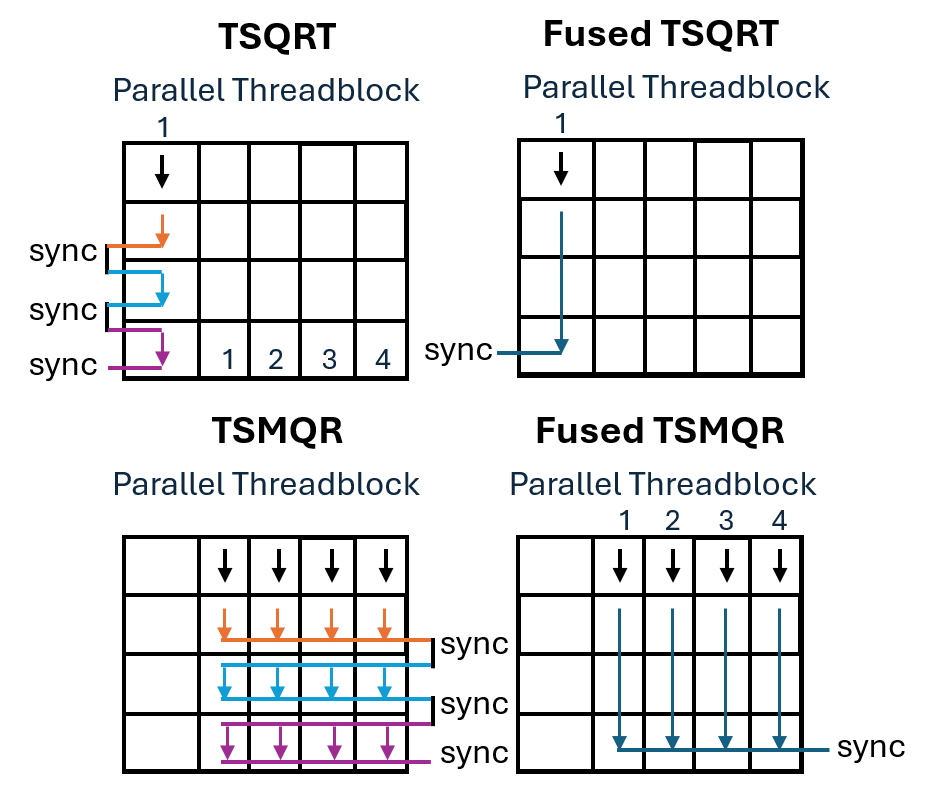}
  \caption{Fused TSQRT kernels (top left) and TSMQR kernels (bottom left) have one single kernel process all rows, while a traditional row-by-row processing of TSQRT kernels (top right) or TSMQR kernels (bottom right) launches a separate kernel for every row and synchronization occurs between every kernel launch. Every arrow represents a workgroup/thread block processing a data tile. The top row (black arrow) gets loaded from global memory for every tile row in the classic TSQRT and TSMQR, but only once for the fused kernel.}\label{fig:fuse}
  \Description{ Visual representation of processing each row by one kernel and synchronizing in between versus processing all rows by a single kernel.  }
  \vspace{-0.3cm}
\end{figure}

This strategy reduces memory traffic by loading the top tile row only once, rather than reloading it for each tile row, and also minimizes kernel launch overhead. This optimization becomes particularly important for large matrix sizes: the number of kernel launches scales quadratically with matrix size when using unfused kernels, but only linearly with fused kernels.
}

{\paragraph{Implementation.}
Algorithm~\ref{alg:kakernel} illustrates the implementation of the custom parallelization strategy for \texttt{UNMQR} (Algorithm~\ref{alg:unmqr}), extended to support \texttt{TSQRT} by including a second tile (denoted $Y$) and fused across rows, as described in the previous paragraph. This fused kernel, referred to as \texttt{TSMQR}, is implemented using the Julia \texttt{KernelAbstractions.jl} framework~\cite{kajl}.  The code is shown with only minor simplifications for readability, such as replacing global index calculations with ellipses (\texttt{...}). Lines 14--23 in Algorithm~\ref{alg:kakernel}, which handle the loading of global data into block or thread memory, correspond to lines 5--8 in Algorithm~\ref{alg:unmqr}. Conversely, lines 36--48 in Algorithm~\ref{alg:kakernel}, responsible for writing data back to global memory, correspond to line 15 in Algorithm~\ref{alg:unmqr}. The application of the Householder vector (lines 25--33) corresponds directly to lines 10--12 in Algorithm~\ref{alg:unmqr}.
}

{\begin{table}[ht]
\small
\centering
\caption{Relative error between the unified implementation (in brackets the cuSOLVER implementation) and the solution, demonstrating accuracy. Maximum error of 30 different matrices with an arithmetic, logarithmic, and semi-circle singular value distribution. }\label{tab:relerror}

\begin{tabular}{|r|c|c|c|}

\hline
\textbf{Matrix size}  & \multicolumn{3}{|c|}{\textbf{Relative error:  unified (cuSOLVER)}}  \\ 
\textbf{ ($n \times n$) } & \textbf{ FP64  }            & \textbf{ FP32}              & \textbf{ FP16 }    \\ \hline
64       & 5.8$ E{-16}$ (5.6$ E{-16}$) & 9.6$ E{-8}$ (5.7$ E{-8}$) & 4.3$ E{-3}$ \\ 
256    & 8.3$E{-16}$ (8.1$E{-16}$) & 8.1$E{-8}$ (4.8$E{-8}$) & 3.3$E{-3}$ \\ 
1024  & 1.4$E{-15}$ (1.6$E{-15}$) & 7.2$E{-8}$ (3.6$E{-8}$) & 6.4$E{-3}$ \\ 
4096   & 3.7$E{-15}$ (6.1$E{-15}$) & 6.7$E{-8}$ (3.5$E{-8}$) & 6.2$E{-3}$ \\ 
16384  &{\color{white}} 6.1$E{-15}$ (1.7$E{-14}$) {\color{white}}& {\color{white}}8.7$E{-8}$ (4.5$E{-8}$) {\color{white}}&{\color{white}} 9.7$E{-3}$ {\color{white}}\\ \hline

\end{tabular}
\vspace{-0.2cm}

\end{table}

\begin{table*}[ht]
\footnotesize
 \caption{Hardware used for benchmarking in Section \ref{res}.}  \label{tab:hardware}
 \vspace{-0.2cm}
  \label{tab:freq}
  \begin{tabular}{cccccl}
    \toprule
    \textbf{CPU (Clock speed) } &  \textbf{GPU (Memory)} &  \textbf{ GPU Multiprocessors} & \textbf{ Memory bandwidth}  &  \textbf{Peak FP32 TFLOPS}  \\
    &&  \textbf{ (L1 memory; L2 memory)}&& \textbf{(Boost clock speed MHz)}\\
    \toprule
    Intel Xeon Platinum 8462Y (2.8GHz) & NVIDIA H100 (80GB) & 132 (256KB; 50 MB) & 3.36TB/s & 67 TFLOPS (1980MHz)  \\
   \midrule
     Intel Xeon Gold 6330 (2.0GHz) & NVIDIA A100 (80GB) & 108 (192 KB; 80 MB) &1.94TB/s & 19.5 TFLOPS (1410MHz) \\
    \midrule
    Intel Core i7-14650HX (2.20GHz) & NVIDIA RTX4060 (8GB)  & 24 (128b KB; 96 MB) & 	272 MB/s & 15.1 TFLOPS (2125MHz) & \\
    \midrule
    AMD Trento EPYC 7A53 (2GHz) & AMD MI250  (128GB) & 208 (16KB; 16MB) &3.28 TB/s & 45.3 TFLOPS (1700 MHz)\\
    \midrule
    MacBook Pro (2.064GHz) & Apple M1 Pro (16GB) & 8 & N.A. & N.A. \\
    \midrule
    Intel Xeon Max 9470C (2GHz) & Intel Ponte Vecchio (64GB) & 1024 (64KB; 408MB) & 3.28 TB/s & 52.4 TFLOPS (1600MHz) \\

  \bottomrule
\end{tabular}
\end{table*}

\paragraph{Accuracy.}
The Householder QR-based SVD is known to be backward stable, with a normwise backward error bounded by a constant times the square root of the matrix size multiplied by machine epsilon~\cite{doi:10.1137/22M1514817}. We confirm numerically that this implementation maintains high accuracy on typical dense matrices by constructing test matrices of the form $A = U^\prime \Sigma V$, where $\Sigma$ contains known singular values, and $U$, $V$ are random unitary matrices~\cite{randommatrices}. We evaluate three singular value distributions over the interval $[0,1]$: an arithmetic distribution, a logarithmic distribution, and a quarter-circle distribution. The first is expected to yield the highest accuracy due to even spacing of singular values; the second is more representative of typical practical cases; and the third mimics the expected spectrum of square matrices with i.i.d. random entries. The $[0,1]$ interval is generalizable, as matrices with larger singular values can be scaled element-wise prior to computation. For each matrix size and singular value distribution, ten test matrices are generated in half, single, and double precision. Table~\ref{tab:relerror} reports the maximum relative Frobenius norm errors across 10 runs, along with those of the reference implementation, \texttt{cuSOLVER}. The results confirm the accuracy of our implementation across precisions and problem sizes, with alignment to the reference solver. No precision-specific techniques, such as rescaling, are applied in the current implementation. Future work may include default rescaling for matrices with singular values outside the target precision range, as well as mixed-precision strategies to enhance accuracy.}

\subsection{Finetuning of Hyperparameters}\label{sect:param}
While a unified API enables the development of a single algorithmic implementation across multiple hardware types and data precisions, hardware-specific constraints inevitably raise questions about performance portability. For example, the amount of data that fits in a cache line differs between half and double precision, and while NVIDIA warps consist of 32 threads, AMD wavefronts use 64. Despite these architectural differences, as shown in the benchmarking results that follow, achieving performance portability does not require complete reimplementation for each hardware and precision configuration. A more development-efficient alternative is to abstract hardware-sensitive parameters, fine-tune them by use case, and leverage Julia’s type inference and multiple dispatch to automatically select optimized configurations based on input types at compile time. 
\begin{table}[ht]
\centering
\small
\caption{Hyperparamter performance tuning is critical. Increasing the COLPERBLOCK improves performance for all architectures and precision, while increasing TILESIZE improves performance at large matrix sizes for three out of four cases. The reference parameters are \texttt{SPLITK}=8, \texttt{TILESIZE=32}, \texttt{COLPERBLOCK=32} and a single parameter is varied. }\label{tab:params}

\begin{tabular}{|r|c|c|c|c|}
\hline
                                                 & \multicolumn{2}{c|}{\textbf{H100}}                                      & \multicolumn{2}{c|}{\textbf{MI250}}              \\ \hline
\textbf{}                                        & \textbf{FP32} & \textbf{FP64} & \textbf{FP32}  & \textbf{FP64} \\ \hline
\multicolumn{5}{|c|}{\textbf{TILESIZE 64 to 32}}               \\ \hline
\textbf{128}                                     & 38\%          & 39\%          & 30\%           & 30\%          \\ \hline
\textbf{512}                                     & 40\%          & 41\%          & 32\%           & 38\%          \\ \hline
\textbf{2048}                                    & 23\%          & 23\%          & 15\%           & 35\%          \\ \hline
\textbf{8192}                                    & 2\%           & 1\%           & -10\%          & 37\%          \\ \hline
\textbf{32768}                                   & -12\%         & -7\%          & -21\%          & 50\%          \\ \hline
\multicolumn{5}{|c|}{\textbf{COLPERBLOCK 32 to 16}}       \\ \hline
\textbf{128}                                     & 2.1\%         & 0.0\%         & 0.0\%          & -1.0\%        \\ \hline
\textbf{512}                                     & 0.7\%         & 0.0\%         & -0.2\%         & 0.0\%         \\ \hline
\textbf{2048}                                    & 0.6\%         & 0.5\%         & 0.0\%          & -0.1\%        \\ \hline
\textbf{8192}                                    & -0.1\%        & 0.1\%         & -4.1\%         & -7.1\%        \\ \hline
\textbf{32768}                                   & -3.6\%        & -9.9\%        & -21.1\%        & -38.2\%       \\ \hline
\end{tabular}
\vspace{-0.3cm}
\end{table}

{In this work, we consider one algorithmic parameter, \texttt{TILESIZE}, and two computational parameters: \texttt{COLPERBLOCK}, which affects the trailing submatrix update, and \texttt{SPLITK}, which impacts the panel factorization. For each architecture and precision type, a brute-force hyperparameter search was conducted to identify optimal values. We tested \texttt{TILESIZE} values between 4 and 128. Since the tile is kept in registers, the product \texttt{TILESIZE} $\times$ \texttt{TILESIZE} $\times$ \texttt{sizeof} \texttt{(precision)} must fit within the available L1 shared memory (SM) for optimal performance. Similarly, \texttt{COLPERBLOCK} is constrained by available register space. For \texttt{SPLITK}, we explored values from 1 up to \texttt{min(TILESIZE, 1024/TILESIZE)}, considering the thread block size limitations. While a full study correlating performance with architectural characteristics is deferred to future work, we highlight the impact of tuning two parameters—\texttt{TILESIZE} and \texttt{COLPERBLOCK}—in Table~\ref{tab:params}. We observe up to a 50\% performance difference when varying a \textbf{ single} parameter while holding others fixed. In particular, \textbf{ larger} \texttt{TILESIZE} values tend to benefit NVIDIA GPUs and AMD GPUs in single precision for \textbf{larger} matrix sizes, whereas they degrade performance on \textbf{smaller} matrices due to reduced occupancy. On AMD GPUs with double precision, \textbf{ smaller} \texttt{TILESIZE} values perform better, likely due to smaller L1 caches compared to NVIDIA, and larger memory occupied relative to single precision. Increasing \texttt{COLPERBLOCK} consistently improves performance across all architectures and precisions, possibly by reducing warp divergence or enhancing memory coalescing through cooperative loading of Householder vectors by threads in a block. The performance impact of \texttt{COLPERBLOCK} is more pronounced on AMD GPUs, likely due to differences in warp execution and memory subsystem design.

These results support the conclusion that unified, portable algorithmic implementations are feasible—but only when hyperparameters are adaptable to hardware and precision. This aligns with findings from prior work on GPU portability~\cite{10.1007/978-3-031-69577-3_7}.

\begin{figure*}[ht]
  \centering
  \includegraphics[width=\textwidth]{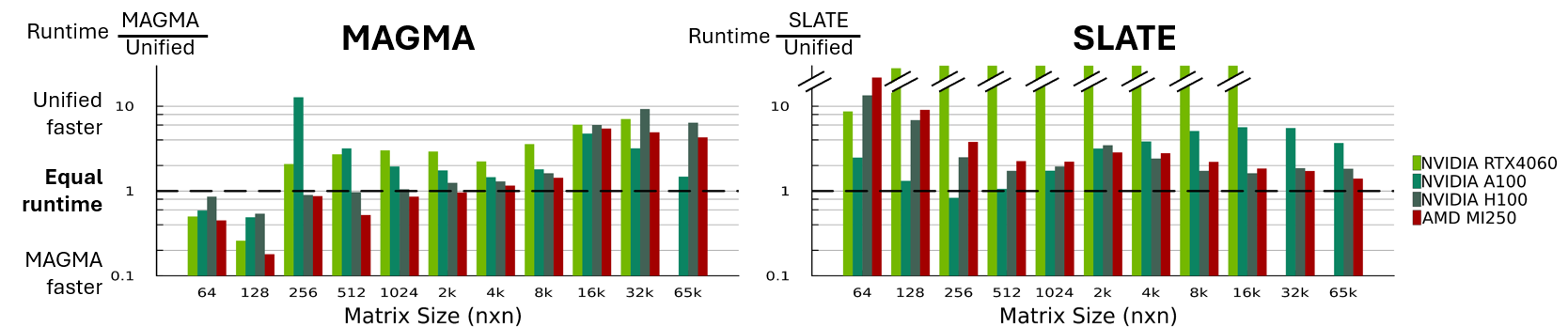}
  \caption{The runtime ratio of singular value calculation in the unified API to state-of-the-art libraries, with higher values indicating that the unified function is faster. The unified API outperforms SLATE for all data sizes and MAGMA (left) for input data larger than 1024. RTX4060 is limited to 32k due to memory size. }
  \Description{Two bar charts displaying the runtime ratio of the unified implementation with reference implementations.}\label{fig:slatemagma}
\end{figure*}
\subsection{Benchmarking}\label{sect:benchmark}
The unified API and vendor-optimized algorithms are benchmarked over 20 runs with a single synchronization at the end to obtain consistent measurements\cite{benchmark}, and this is repeated until 2 seconds total benchmark time is reached. {The \texttt{cuSOLVER}/\texttt{rocSOLVER} benchmarks (\verb|rocsolver_Xgesvd|, \verb|cusolverDnXgesvd|) were verified in C++ to produce identical performance measurements. The Intel implementation was executed via \verb|oneapi::mkl::lapack::gesvd|. For comparison, \texttt{MAGMA} (v2.9.0) \verb|testing_Xgesvd| and \texttt{SLATE} (v2024.10.29) \verb|svd| were run from the command line using their respective tester executables.} MAGMA is run with 1 GPU and no singular vectors specified. SLATE is run with as options the target and origin being the device, and does not calculate singular vectors by default. Each measurement is run twice after each other to obtain more consistent measurements. The hardware used for benchmarking is specified in Table~\ref{tab:hardware}. Apple Metal's technical specifications are not readily published by Apple.

\begin{figure}[ht]
  \centering
  \includegraphics[width=\linewidth]{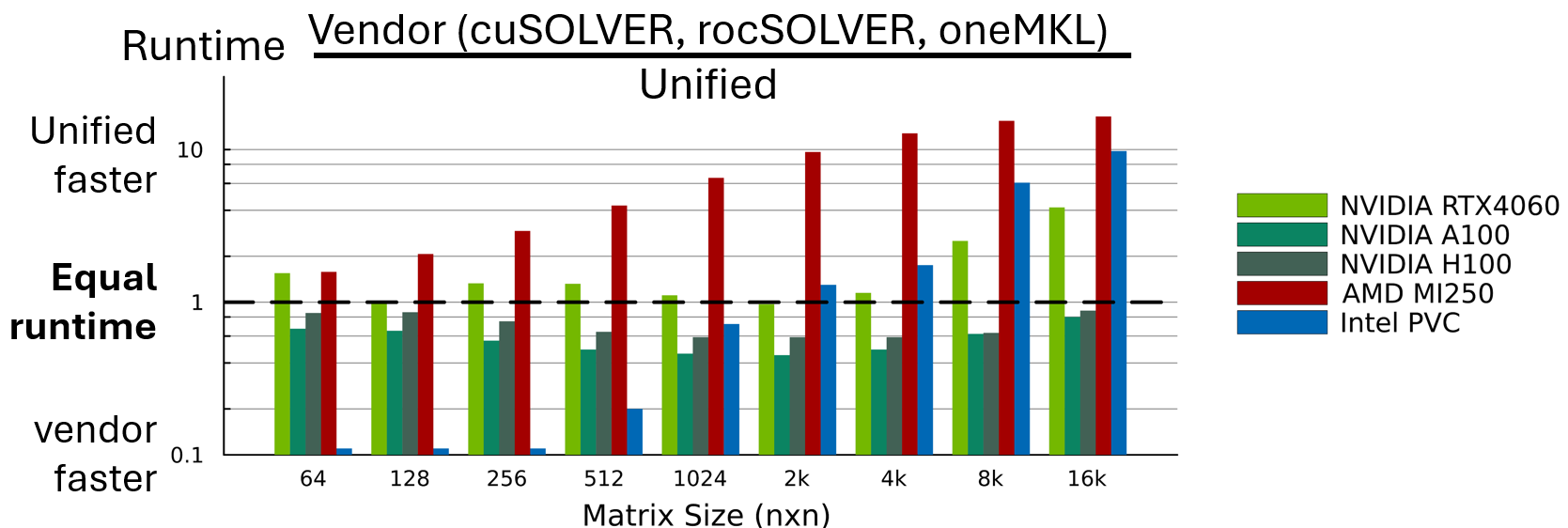}
  \caption{The runtime ratio in the unified function to vendor libraries, with higher values indicating that the unified function is faster. The unified function outperforms rocSOLVER (MI250) for all data sizes, cuSOLVER on consumer GPUs (RTX4060), and Intel OneMKL at matrix sizes larger than 2048. It approaches cuSOLVER on HPC architectures (A100 and H100) at 80-90\% at large matrix sizes, and at 50-80\% for smaller data sizes. }
  \Description{Bar charts displaying the runtime ratio of the unified implementation with vendor-provided reference implementations.}\label{fig:curoc}
\end{figure}

\begin{figure*}[ht]
  \centering
  \includegraphics[width=\textwidth]{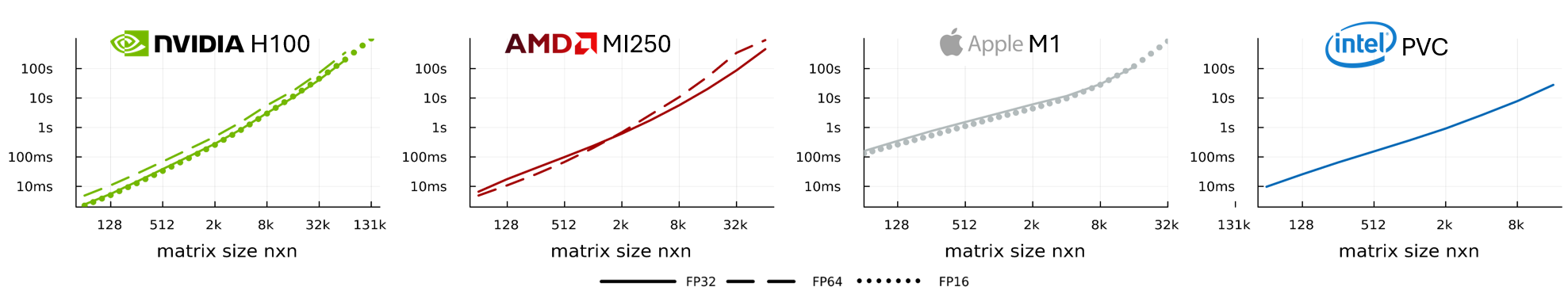}
  \caption{Portability across precision and hardware: the runtime of the unified function for singular values for NVIDIA H100, AMD MI250, Apple M1, and Intel PVC for FP16, FP32. On NVIDIA, FP16 has the same speed as FP32 because it uses the FP32 CUDA cores, but goes up to larger data sizes. Julia AMD GPU  currently does not support conversion at calculation time for FP16, and Apple Metal does not support FP64. Only single precision results are shown for Intel GPUs, based on a test case we provided to Intel collaborators for evaluation.}\label{fig:crosshardware}
  \Description{Four plots of runtime in function of matrix size with logarithmic axis, showing a comparable linear relationship for each hardware architecture.}
\end{figure*}

\section{Performance Results}\label{res}
To demonstrate unified performance and portability, we discuss performance versus state-of-the-art linear algebra libraries in Section \ref{sect:perf}, performance across hardware vendors and data precision in Section \ref{sect:port}, and hardware-specific hyperparameter optimization in Section \ref{sect:param}.
\subsection{Performance Relative to State-of-the-Art Linear Algebra Libraries}\label{sect:perf}

Figure \ref{fig:slatemagma} displays the relative runtime of the unified implementation over the state-of-the-art optimized libraries MAGMA (left) and SLATE (right), and Table \ref{tab:summaryfigures} summarizes the values in the Figure.
We can see in the right chart that the unified implementation exceeds the speed of SLATE for both AMD (MI250) and NVIDIA hardware, and for all types of NVIDIA hardware: consumer GPUs (RTX4060), older generations of high-performance GPUs (A100), and the newest generation of high-performance GPUs (H100). In the left bar chart, we see the unified implementation exceeding equal runtime for all hardware on matrix sizes larger than 2048 $\times$2048. For smaller matrix sizes between 256 and 1024, the unified implementation performs on par with the MAGMA library, while for matrices below 256, we see a moderate performance penalty (20-90\% of the MAGMA performance is reached). The implemented unified kernels are optimized for large matrix sizes; therefore, a modest reduction in performance for smaller matrices is expected. In addition, the runtime of matrices smaller than $ 256\times256$ matrix are less than 40 milliseconds, making performance optimization tedious. In Section \ref{sect:param}, we will discuss hyperparameter tuning for performance and remark that if we wish to improve performance for smaller matrix sizes in the future, different hyperparameter values by matrix size could contribute to faster performance. To conclude, our results show that for matrix sizes larger than 256, the unified API matches or surpasses the performance of MAGMA, and outperforms SLATE across the full range of matrix sizes. {The \texttt{MAGMA} and \texttt{SLATE} libraries are designed for large-scale problems that leverage multi-GPU and hybrid CPU--GPU systems, which may explain \texttt{SLATE}'s lower performance on consumer-grade laptops. Moreover, both libraries rely on vendor-optimized backends such as \texttt{cuSOLVER} and \texttt{rocSOLVER} for low-level kernels, requiring vendor-specific implementations. In contrast, our general-purpose, unified singular value decomposition function achieves high performance across a wide range of platforms—from consumer laptops to HPC clusters with any GPU hardware—without changing a single line of code. This portability enables seamless integration into diverse environments. In the future, large libraries such as \texttt{SLATE} or \texttt{MAGMA} could adopt our open-source solution as a drop-in replacement for vendor-specific components at the low level.
}
\begin{table}[ht]
\centering
\small
\caption{Geometric mean of singular value runtime ratios of unified API to vendor and HPC libraries over matrix sizes in Figures \ref{fig:slatemagma}-\ref{fig:curoc}. (Range between brackets) }\label{tab:summaryfigures}

\begin{tabular}{|r|c|c|c|c|c|}

\hline
                        & \textbf{vendor}  & \textbf{MAGMA}  & \textbf{SLATE}  \\ \hline
\textbf{NVIDIA RTX4060} & 1.5 (1.0 - 4.2)  & 2.2 (0.3 - 7.1) & 280 (9 - 2200)  \\ \hline
\textbf{NVIDIA A100}    & 0.6 (0.5 - 0.8)  & 2.1 (0.5 - 13)  & 2.5 (3.2 - 5.7) \\ \hline
\textbf{NVIDIA H100}    & 0.7 (0.6 - 0.9)  & 1.5 (0.5 - 9.3) & 2.8 (1.6 - 13)  \\ \hline
\textbf{AMD MI250}      & 5.9 (1.6 - 16)   & 1.0 (0.2 - 5.5) & 3.4 (1.7 - 22)  \\ \hline
\textbf{Intel PVC}      & 0.5 (0.03 - 9.8) & -               & -               \\ \hline
\end{tabular}
\end{table}
Figure~\ref{fig:curoc} displays the relative runtime compared to the vendor-optimized libraries. The matrix sizes in this bar chart only go up to 16k, as the current version of NVIDIA's cuSOLVER (12.8) suffers from an issue in supporting 64-bit addressing for eigensolvers and singular value solvers~\cite{eigissuecuda}, and AMD's rocSOLVER has not implemented 64-bit addressing support for these functions yet~\cite{rocsolver}. The bar chart shows the unified API exceeding AMD's rocSOLVER on MI250 architecture for all matrix sizes. On consumer GPUs (RTX4060), the unified implementation also outperforms cuSOLVER for all data sizes.
On high-performance NVIDIA GPUs (A100 and H100), the unified API achieves 80-90\% of the performance of cuSOLVER for large matrix sizes of 8k and 16k and 50-90\% of its performance for other matrix sizes. {The difference between consumer and HPC GPUs could be caused by the differing characteristics: HPC GPUs have a larger number of SMs, but can have lower clock speeds than high-range consumer GPUs (Table \ref{tab:hardware}). }  On Intel, the unified API exceeds the performance of oneMKL's native implementation for matrix sizes over 2048. Similarly, the kernels presented in this work are optimized for large matrix sizes, and the underperformance observed at smaller sizes is expected. This could be addressed in future work through matrix size–specific hyperparameter tuning. We conclude that the unified API offers a performance advantage over vendor-optimized implementations on AMD GPUs, on NVIDIA consumer hardware, and for large matrix sizes on Intel GPUs. On high-end NVIDIA GPUs, the unified API achieves 50–90\% of cuSOLVER’s performance.

{\subsection{Relative Runtime of Individual Kernels}
To understand the bottlenecks of the algorithm, we analyze the relative runtime of the different computational stages, as shown in Figure~\ref{fig:subfunc}. Stage one, which includes panel factorization and trailing submatrix update during the reduction to band form, operates on a substantially larger portion of the matrix than the subsequent reduction to diagonal form, particularly as the matrix size increases. This naturally makes stage one more time-consuming in relative terms
{at increasing matrix size}: the trailing submatrix update affects a proportionally larger part of the matrix as size increases, whereas the panel factorization always processes a single panel. 
{While the trailing matrix update processes more data overall, it can achieve similar—or even better—performance than the panel factorization at small matrix sizes due to superior parallelization and GPU occupancy. The panel factorization operates using a single thread block and involves more extensive serial computation, whereas the trailing update is a highly parallel matrix-multiply-like operation that can reach full occupancy. However, once full occupancy is surpassed, the relative runtime of the trailing update increases more rapidly than that of the panel factorization due to its larger data footprint. Beyond this point, additional computations are no longer executed in parallel across GPU multiprocessors but are instead serialized: more thread blocks are run per GPU multiprocessor than its maximum parallel capacity. On GPUs with fewer Streaming Multiprocessors (SMs), such as the RTX 4060, full occupancy is reached at smaller matrix sizes. This effect is evident in the steep increase in the trailing update–to–panel factorization runtime ratio observed between matrix sizes of 8k and 32k.
}
 AMD GPUs require a proportionally longer time for the first stage compared to NVIDIA GPUs, which may be attributed to differences in peak FLOP throughput and register file sizes—the former impacting the compute-bound panel factorization, and the latter affecting the memory-bound trailing update. We conclude that for large matrix sizes, the primary bottleneck lies in the trailing submatrix update, whose optimization strategy is detailed in Section~\ref{sect:gpukernel}.
\begin{figure}[ht]
  \centering
  \includegraphics[width=\linewidth]{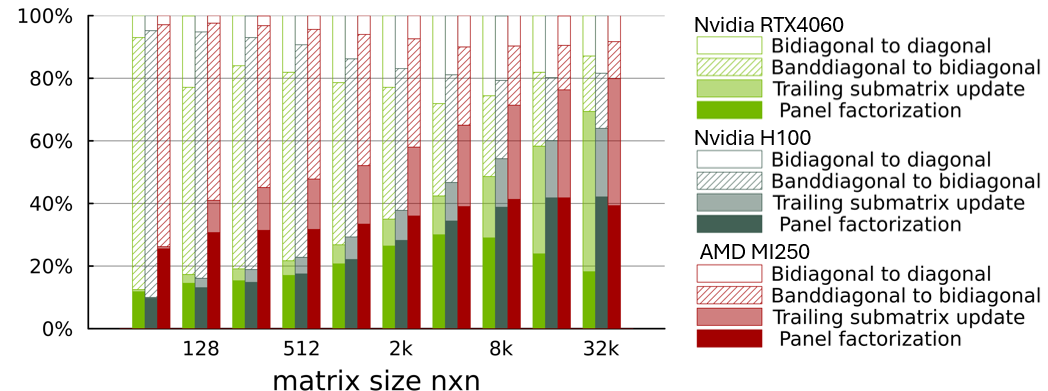}
  \caption{The relative runtime of the panel factorization, trailing submatrix update, reduction to bidiagonal, and reduction to diagonal. As the matrix size increases, the first stage (reduction to band form) becomes more important. Additionally, as the matrix size increases, the ratio of trailing submatrix time to panel factorization time also increases.}
  \Description{Bar charts displaying the runtime ratio of subkernels for matrix sizes 64 to 32k in percentage. }\label{fig:subfunc}
\end{figure}

}
\subsection{Portability Across Hardware and Precision}\label{sect:port}
Figure~\ref{fig:crosshardware} shows the runtime of the unified implementation across several hardware backends: NVIDIA H100, AMD MI250, Apple Metal M3, and Intel One Max GPU for FP64, FP32, and FP16 data types. The plot demonstrates the portability of the API across hardware and precision. Note that for each hardware and data type, the optimal hyperparameter combination was selected. In the NVIDIA plot, we observe that the single and half precision performance curves are nearly identical. This is because the H100 GPU does not support scalar FP16 arithmetic on its regular CUDA cores; instead, FP16 operations are designed to run efficiently only on Tensor Cores. In our implementation, Tensor Cores are not explicitly used, so FP16 inputs are upcast to FP32 during computation and downcast at storage time, resulting in performance similar to single precision.
As a result, support for half precision on NVIDIA GPUs through the unified API enables GPU-resident computations for larger matrix sizes (up to $131k$ $\times$ $131k$) than previously possible with single or double precision. On Apple Metal, we observe performance trends similar to NVIDIA. This work presents the first implementation of a GPU-accelerated singular value solver on Apple GPUs, supporting both single and half precision. 
\section{Conclusion}

In this work, we implemented GPU kernels for computing singular values using QR iteration and demonstrated the viability of a unified API across both data precisions and hardware backends, leveraging Julia’s abstraction frameworks, including \texttt{GPUArrays} and \texttt{KernelAbstractions}. We achieved portability across NVIDIA, AMD, Apple Metal, and Intel GPUs, and across half precision (FP16), single precision (FP32), and double precision (FP64). In terms of performance, the unified API outperforms \texttt{rocSOLVER}, performs on par with cuSOLVER (achieving 80–90\% of its speed for large matrices) on high-end GPUs, outperforms cuSOLVER on consumer GPUs, and surpasses MAGMA and SLATE for matrices larger than 1024. To our knowledge, this work presents the first GPU-accelerated singular value solver for Apple Metal and the first to support half precision (FP16) on GPUs. While most current GPUs support FP16 only through tensor cores and not scalar arithmetic units, our implementation is designed to be forward-compatible with future hardware that includes scalar FP16 support. This is especially relevant for emerging AI workloads such as large language models (LLMs) with low-rank adaptation (LoRA), which increasingly rely on low-precision arithmetic. Finally, we emphasized the importance of hyperparameter tuning to achieve performance portability within unified APIs.
Building on this foundation, future work includes a detailed performance breakdown and benchmarking of each stage of the singular value computation pipeline, 
as well as a function-by-function comparison with open-source reference software (i.e., rocSOLVER, MAGMA), and a detailed analysis of the performance variation across different hardware backends of one vendor. Function-by-function comparison with OneMKL or cuSOLVER is not possible as the software is proprietary.
We plan to extend the implementation to compute singular vectors, enabling full-rank SVD functionality. Further work will also focus on automated tuning of kernel parameters, adapting to input matrix sizes and data types for improved performance on small and irregular workloads. In addition, we aim to incorporate support for out-of-core execution, multi-GPU scaling, and heterogeneous environments, enabling larger problem sizes and better resource utilization. Support for non-square matrices and specialized algorithms for tall and skinny matrices are also the subject of further work. Finally, we envision integrating the solver into distributed task-based frameworks, such as \texttt{Dagger}~\cite{alomairy2024dynamic}, to facilitate dynamic scheduling, pipelining, and composability in larger scientific applications.

{

\begin{acks}

We would like to acknowledge the work and support of members of the Julia Lab, in particular James Schloss, Julian Samaroo, and Tim Besard, who helped with our understanding of available tools within the Julia ecosystem for performance portability. We would also like to thank Huda Ibeid for benchmarking the implementation on Intel systems.
We acknowledge the Belgian American Educational Foundation for supporting Evelyne Ringoot and the KAUST Ibn Rushd post-doctoral fellowship for supporting Rabab Alomairy. We would like to acknowledge IBEX at the Supercomputing Laboratory of the King Abdullah University of Science and Technology (KAUST) in Thuwal, Saudi Arabia, and the MIT Office of Research Computing and Data for computational resources.
This material is based upon work supported by the US NSF (CNS-2346520, PHY-2028125, RISE-2425761, DMS-2325184, OAC-2103804), DARPA (HR00112490488), DoE (DE-NA0003965), and USAFR (FA8750-19-2-1000). Neither the United States Government nor any agency thereof, nor any of their employees, makes any warranty, express or implied, endorses, recommends, or favors, or assumes any legal liability or responsibility for the accuracy, completeness, or usefulness of anything in this report, or represents that its use would not infringe privately owned rights. The views and opinions expressed herein are those of the authors and do not necessarily state or reflect those of the U.S. Government or any agency thereof, nor constitute official policies, either expressed or implied, of the U.S. Government or any agency thereof.

\end{acks}
}

\bibliographystyle{ACM-Reference-Format}
\bibliography{sample-base.bib}

\end{document}